\documentclass[]{aastex62}

\received{December 3, 2018}
\revised{January 10, 2019}
\accepted{January 10, 2019}
\submitjournal{ApJL}

\shorttitle{Large Volcanic Event on Io}
\shortauthors{Morgenthaler et al.}

\begin{document}

\title{\replaced{Large Volcanic Event on Io Detected During 2018
    Jovian Opposition by PSI's Io Input/Output Facility (IoIO)}{Large
    Volcanic Event on Io Inferred from Jovian Sodium Nebula Brightening}}

\correspondingauthor{Jeffrey P.\ Morgenthaler}
\email{jpmorgen@psi.edu}

\author[0000-0003-3716-3455]{Jeffrey P.\ Morgenthaler}
\affiliation{Planetary Science Institute \\
  1700 East Fort Lowell, Suite 106\\
  Tucson, AZ 85719-2395, USA}

\author[0000-0001-7619-652X]{Julie A.\ Rathbun}
\affiliation{Planetary Science Institute \\
  1700 East Fort Lowell, Suite 106\\
  Tucson, AZ 85719-2395, USA}

\author[0000-0002-6917-3458]{Carl A.\ Schmidt}
\affiliation{Center for Space Physics\\
  Boston University\\
  Boston, MA 02155, USA}

\author[0000-0003-2310-6628]{Jeffrey Baumgardner}
\affiliation{Center for Space Physics\\
  Boston University\\
  Boston, MA 02155, USA}

\author[0000-0001-6720-5519]{Nicholas M.\ Schneider}
\affiliation{University Of Colorado, Boulder\\
  Boulder, CO 80309, USA}



\begin{abstract}

  \deleted{Using narrow-band images of the Jovian sodium nebula we
    detected a large release of gas from Io, gas which we assume to
    originate from a volcanic event.  The images were recorded on over
    150 nights by the 35\,cm coronagraph which comprises PSI's Io
    Input/Output Facility (IoIO).  The onset of the event occurred in
    the mid December 2017 -- early January 2018 timeframe.  Sodium
    emission over the IoIO 0.4\mbox{$^{\circ}$} field-of-view of was
    seen to increase through January 2018 and peak in early March
    2018.  By early June 2018, the surface brightness of the emission
    returned to the value seen 2017 April -- June, making this the
    longest such event observed by this technique \citep{brown97} and
    comparable in length to that observed by the \textsl{Galileo} Dust
    Detector in 2000 \citep{krueger03}.  No other report of this event
    has been made despite a significant number of observations of the
    Jovian system by and in support of NASA's \textsl{Juno} mission.
    This detection therefore places those observations in valuable
    context and highlights the importance of synoptic observations by
    facilities such as IoIO, which provide a global view of neutral
    material in the Jovian magnetosphere.}

\added{Using narrow-band images recorded on over 150 nights by the
  35\,cm coronagraph which comprises PSI's Io Input/Output Facility
  (IoIO), we detected a 6-month long enhancement in the Jovian sodium
  nebula.  The onset of the enhancement occurred in the mid December
  2017 -- early January 2018 timeframe.  Sodium emission over the IoIO
  0.4\mbox{$^{\circ}$} field-of-view of was seen to increase through
  January 2018 and peak in early March 2018.  By early June 2018, the
  surface brightness of the emission returned to the value seen 2017
  April -- June, making this the longest such event observed by this
  technique \citep{brown97, yoneda15} and comparable in length to that
  observed by the \textsl{Galileo} Dust Detector in 2000
  \citep{krueger03}.  A new IR hot-spot was found on Io near
  Susanoo/Mulungu paterae between January 2 and 12, however this
  hot-spot was neither bright nor long-lasting enough to have been
  independently identified as the source of a major sodium nebula
  enhancement.  Furthermore, no other report of this event has been
  made despite a significant number of observations of the Jovian
  system by and in support of NASA's \textsl{Juno} mission.  This
  detection therefore places those observations in valuable context
  and highlights the importance of synoptic observations by facilities
  such as IoIO, which provide a global view of neutral material in the
  Jovian magnetosphere.}
  
\end{abstract}



\section{Introduction}
\label{intro}

Io's volcanism was first hinted at by a fortuitous observation in the
3 -- 5\,$\mu$m region of the infrared \citep{witteborn79}, though it
was not understood as such until after \textsl{Voyager}~1 observations
confirmed the presence of plumes \citep{morabito79, sinton80b}.  This
volcanism helped to place in context earlier fortuitous observations
of Io's ionosphere \citep{kliore75}, a sodium cloud near Io
\citep{brown74}, and ionized sulfur emission near Jupiter
\citep{kupo76}: Io has an atmosphere which ultimately derives its
source from volcanic activity and supplies Jupiter's magnetosphere
with a substantial amount of material \citep[$\sim$1\,ton\,s$^{-1}$,
  e.g.,][]{mcgrath04, schneider07}.  As discussed in these references,
material that is ionized forms the Io plasma torus (IPT), which
encircles Jupiter near Io's orbital radius.  It is the bright line of
singly ionized sulfur at [S$\;${\small\rmfamily II}]\,6731\,\AA\ which
led to the initial detection of the IPT and has enabled it to be
imaged by ground-based coronagraphs with apertures as small as 30\,cm
\citep{nozawa04_no_SII}.  Interaction between the IPT and Io's
atmosphere via processes such as sputtering, charge exchanging and
dissociative recombination, result in the energetic ejection of
neutral material.  Although a minor component of the material that is
released, sodium has such bright doublet emission at 5890\,\AA\ and
5896\,\AA, that it has been imaged by ground-based coronagraphs with
apertures as small as 10\,cm \citep[e.g.,][]{mendillo90, mendillo04,
  yoneda09, yoneda10, yoneda14, yoneda15}.

\citet{mendillo04} used of order one wide-field (6\mbox{$^{\circ}$})
sodium cloud image per year between 1990 and 1998 and a literature
search of available Io infrared measurements to suggest there was a
general correlation between the sodium nebula brightness and Io's
disk-averaged infrared brightness (their Figure~2).  Long-lived
volcanic hot spots, particularly Loki Pathera and Tiermes Pathera,
were identified as the primary causes of this correlation (their
Figure~1).  Subsequent work by \citet*{deKleer16AGU} \explain{expanded
  author list to 3 to make clear differentiation with new reference
  \citep{deKleer16}} used 3-years of higher cadence sodium cloud
images (up to one per day) and much higher spatial resolution IR
monitoring and failed to confirm this correlation.  Rather,
\citet*{deKleer16AGU} suggested some, but not all, bright transient IR
events traceable to individual volcanic eruptions may trigger sodium
cloud brightening.  Loki Pathera and Tiermes Pathera are lava lakes,
which are not known to produce high eruptive plumes
\citep[e.g.,][]{rathbun06, depater17}.  Rather, explosive events
produced by volcanoes such as Pele, Tvashtar and Pillan
\citep[e.g.,][]{jessup12} would seem more likely to result in the
ejection of material, though it is not clear if plume material from
these eruptions can be ejected directly beyond Io's atmosphere or if
sublimation of the large ejecta blankets observed around these
volcanoes is responsible for increase in ejection rates.  Finally,
\citet{johnson95} have suggested that SO$_2$ geysers may create
``stealth plumes,'' undetected by methods that monitor Io surface or
near-surface properties, since they would not have strong infrared or
dust signals.  Regardless of the precise physical mechanism operating,
Io's volcanic nature is ultimately responsible for the release of gas
into Jupiter's magnetosphere.  Therefore, for the purposes of this
work, we will call such a release of gas a volcanic event.

Using a spectroscopic study that lasted an entire Jovian opposition,
\citet{brown97} showed that when there was a large increase in sodium
emission in the inner Jovian magnetosphere, the IPT also became
brighter and shifted to the east.  The sodium peak brightness was seen
before the IPT peak brightness.  \citet{brown97} attributed this
behavior to an eruption of a volcanic plume on Io and the resulting
radial and antisunward diffusion of material through the Jovian
magnetosphere.

\citet{yoneda10} used the \citet{nozawa04_no_SII} [S$\;${\small\rmfamily II}]
IPT observations and contemporaneously recorded small-aperture
coronagraphic sodium nebula images to establish a correlation between
IPT brightness and sodium nebula brightness in the same sense as that
found by \citet{brown97}.  Extreme ultraviolet (EUV) observations of
the IPT have also shown evidence of correlation with indicators of
volcanic ejection of material from Io \citep{krueger03, steffl06,
  yoneda15, kimura18}.

Motivated by the success of the small-aperture ground-based
coronagraphic observations of the IPT and Jovian sodium nebula by
\citet{nozawa04_no_SII}, \citet{mendillo04}, \added{and
  \citet{yoneda09, yoneda10, yoneda14, yoneda15}} and the numerous
open scientific questions in inner Jovian magnetospheric studies, we
created the Io Input/Output facility (IoIO).  \added{D}escribed in
more detail in \S\ref{observations}, \added{IoIO is comparable in
  aperture size to the coronagraphs used by \citet{nozawa04_no_SII} so
  that detection of the IPT in [S$\;${\small\rmfamily II}] is
  possible.  This makes the IoIO aperture area $\sim$10-times that of
  the wide-field sodium nebula studies of \citet{mendillo04} and
  \citeauthor{yoneda15}, with a comparable reduction in field-of-view.
  The larger aperture, yet still relatively large field-of-view
  (0.4\arcdeg) simultaneously enables IoIO to study the detailed
  three-dimensional structure of the sodium nebula near Io and Jupiter
  (Figure \ref{fig:Na} and on-line animation) and, like
  \citeauthor{yoneda15}, measure the average surface brightness of the
  Jovian sodium nebula with a nightly cadence.  As detailed in
  \S\ref{observations} -- \S\ref{discussion}, our reduction techniques
  are sufficient to demonstrate the detection of a large and long-term
  enhancement in the sodium nebula, however, removal of the effects of
  passing clouds is not yet as sophisticated as those of
  \citeauthor{yoneda15}.  As a result, the scatter our data is greater
  and the sensitivity to small enhancements is less.  This will be
  addressed in subsequent iterations of our reduction pipeline.}  In
\S\ref{results} -- \S\ref{discussion} and Figure~\ref{fig:Na_vs_T}, we
show that IoIO detected a substantial increase in the amount of sodium
within $\sim$50 Jovian radii ($\mathrm{R_{j}}$) of Jupiter starting in
the mid December 2017 -- early January 2018 timeframe.  In
\S\ref{conclusion}, we suggest this was caused by a volcanic event on
Io.

\section{Observations}
\label{observations}

The Io Input/Output facility (IoIO) consists of a 35\,cm Celestron
telescope feeding a custom-built coronagraph, a boresight-mounted
80\,mm guide telescope and an Astro-Physics 1100 GTO German equatorial
mount.  IoIO is located at the San Pedro Valley Observatory, a hosting
site situated in a dark location 100\,km east of Tucson, Arizona, USA.
The coronagraph imaging system is telecentric: A Kodak Wratten ND3
gelatin neutral density filter cut $\sim$1.5\,mm wide is placed at the
focal plain of the Celestron telescope so that Jupiter is attenuated
rather than occulted, allowing for astrometric and photometric
calibrations.  The diverging light from the f/11 beam then passes
through one of the five filters listed in
Table~\ref{tab:filters}\added{, which are standard bandpasses for this
  work.}  The filters are hard metal oxide coated to maximize
durability and minimize central wavelength (CWL) temperature drift
($<$0.1\,\AA\,C$^{-1}$).  The narrow-band Fabry-P\'erot type filters,
fabricated by Custom Scientific, have a very flat-topped profile with
$>$90\% peak efficiency.  The sodium on-band filter was constructed
such that both the Na D lines are transmitted with $<$1\% change in
efficiency over the entire FOV and nighttime temperature range
expected at our hosting site.  After the filters, the light passes
through a field lens which focuses the telescope pupil onto the pupil
of a Nikon Nikkor 60\,mm F/2.8 camera lens.  Finally, the light is
collected by a Starlight Xpress SX694 medium format CCD camera.  The
effective focal length of IoIO is 1200\,mm, the FOV for sodium nebula
observations is 0.4\mbox{$^{\circ}$}\ or 64\,$\mathrm{R_{j}}$\ --
84\,$\mathrm{R_{j}}$, depending on Jupiter's geocentric distance and
pixels are 0.78\hbox{$^{\prime\prime}$} per side.


\begin{table}
\centering
\caption{Filter Properties$^{a}$}
\label{tab:filters}
\vspace{0.1cm}
\begin{tabular}{lllllllllll}
  \hline
  Filter 	& CWL (\AA)		& FWHM (\AA)\\
  \hline
  R		& 6349			& 1066\\
  {[{S$\;${\small\rmfamily II}}]} on-band& 6731		& 10\\
  Na on-band	& 5893			& 12\\
  {[{S$\;${\small\rmfamily II}}]} off-band& 6640	& 40\\
  Na off-band	& 6000			& 50\\
  \hline
\multicolumn{3}{l}{$^{a}$Measured in a collimated, normal-incidence beam at 20\mbox{$^{\circ}$}\,C}
\end{tabular}
\end{table}

Sodium observations are recorded in on-band/off-band pairs for five
minutes and one minute, respectively, every $\sim$30 minutes.  On-
and off-band images of the Io plasma torus in [S$\;${\small\rmfamily
    II}]\,6731\,\AA\ are recorded on a 6-minute cadence in the
intervals between the Na observations and will be reported in another
work.

\section{Data Reduction}
\label{reduction}

Because the sodium nebula is a field-filling source for our FOV, we
take some care in reducing the data.  This starts with the bias and
dark subtraction of our on- and off-band images.  The IPT is a much
smaller target than the sodium nebula, so as a cost-savings measure,
we used 32\,mm diameter [S$\;${\small\rmfamily II}] on- and off-band
filters compared to the 50\,mm Na filters.  The small
[S$\;${\small\rmfamily II}] filter diameters enabled us to use pixels
on the edges of the [S$\;${\small\rmfamily II}] FOVs, to construct a
near continuous record of the combined effects of bias and dark
current through each night.  These values were interpolated in time
and subtracted from the Na on- and off-band images.

Sky flats show that white light vignetting is $\sim$10\% starting
beyond the region we use for our analyses, so we ignore the effect.
Similarly, we ignore small-scale variation in biases, flats, and darks
since our primary results are derived by averaging over large areas of
pixels.

After bias and dark subtraction, we subtract the off-band image
recorded closest in time from each on-band image.  A factor, OFFSCALE,
is multiplied by each off-band image.  OFFSCALE is the product of the
flux in the central 10\mbox{$\times$}10 pixel
($7.8^{\prime\prime}\times7.8^{\prime\prime}$) areas of Jupiter in the
on- and off-band images times.  An additional factor of 0.80 is
applied to remove over-subtraction consistently seen in the images.
OFFSCALE typically varies by $\sim$20\% each night and there was a
systematic drop of 20\% during April attributable to improvements we
were making in the guiding system: because on-band images have longer
exposure times than off-band, improved guiding reduced smearing of
Jupiter preferentially in the on-band images, hence raising OFFSCALE.
We show in \S\ref{discussion} that the systematic change in OFFSCALE
has no effect on our results.

We derive a factor, ADU2R, to convert pixel values to the surface
brightness unit of rayleighs (R) where 1\,R = ${10^6 \over
  4\pi}$\,photons~s$^{-1}$\,cm$^{-1}$\,sr$^{-1}$:

\begin{equation}
  \mathrm{ADU2R} = {on\_jup * ND \over MR}
  \label{eq:ADU2R}
\end{equation}

\noindent  
Here, $on\_jup$ is the average pixel value of the $10\times10$ pixel
box centered on Jupiter in the on-band images.  This area represents
pixels within $\sim0.2\,\mathrm{R_{j}}$ of the center of Jupiter.  ND
is the attenuation factor provided by our neutral density filter.
R-band measurements of GSC5017:78 on 2018-03-20UT show ND =
$730\pm70$.  $MR$ is the surface brightness of Jupiter over our
12\,\AA-wide bandpass on-band filter.  To account for the deep sodium
Fraunhofer absorption lines, this is calculated using Jovian albedos
from \citet{woodman79} and \citet[see also PDS:
  ESO-J/S/N/U-SPECTROPHOTOMETER-4-V2.0]{karkoschka98} and the
\citet{kurucz05ATLAS} solar flux atlas.  $MR$ varied from 52.6\,MR to
54\,MR over the IoIO observations.

\begin{figure}[h]
\plotone{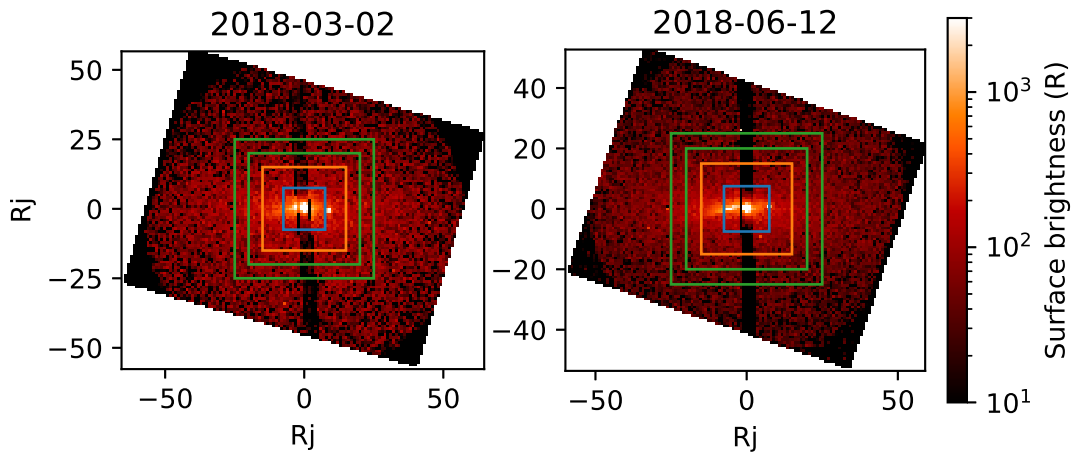}

\caption{Two images of the inner
  0.4\mbox{$^{\circ}$}\ ($\sim$50\,$\mathrm{R_{j}}$) of the Jovian
  sodium nebula recorded by IoIO.  Left: image recorded 2018-02-27
  08:26:11 UT, during the period when the extended Jovian sodium
  nebula was bright.  Right: image recorded 2018-06-12 04:40:37UT,
  during a a period when the nebula was at baseline value.  The images
  have been down-sampled by a factor of four [five for arXiv e-print]
  and then rebinned by another factor of four [five for arXiv
    e-print].  The boxes indicate the apertures used to construct
  Figure~\ref{fig:Na_vs_T} \added{(left)}.  The innermost aperture
  (blue) is a square area 15\,$\mathrm{R_{j}}$ on a side containing
  points approximately within approximately 7.5\,$\mathrm{R_{j}}$ from
  Jupiter (blue triangles in Figure~\ref{fig:Na_vs_T}, \added{left}).
  The next concentric aperture (orange) is a square aperture
  30\,$\mathrm{R_{j}}$ on a side containing points
  $<15\,\mathrm{R_{j}}$ from Jupiter (orange squares in
  Figure~\ref{fig:Na_vs_T}, \added{left}).  The square annular area,
  between the outermost two (green) rectangles was used to calculate
  the surface brightnesses shown as the green Xs in
  Figure~\ref{fig:Na_vs_T} \added{(left and right)}.  This corresponds
  to points $20\,\mathrm{R_{j}} < r < 25\,\mathrm{R_{j}}$, which is
  comparable to the 25\,$\mathrm{R_{j}}$ aperture used by
  \citet{yoneda09}.  An animation\added{, lasting 1.5~min,} of the
  dataset is found in the online Journal \added{and shows the 3-D
    structure of the ``banana,'' ``jet'' and ``stream'' features
    discovered by \citet{schneider91Na} and modeled in detail by
    \citet{wilson02}.}  Individual frames of the animation have been
  processed with the histogram equalization method to enhance contrast
  of low surface brightness features.  Frames with high background
  light (average surface brightness $>$250\,R) have been removed, as
  have frames where the image of Jupiter moved more than 5 pixels
  between the on-band and off-band images.  The later effect does not
  materially affect our aperture surface brightness values, but it
  does detract cosmetically from the animation.  In the animation,
  values above 8\,kR have been set to zero and Jupiter has been scaled
  up by a factor of 100. \added{Long-term, an on-going archive of all
    raw and reduced images collected by IoIO will be kept at NASA's
    Planetary Data System (PDS).}}
\label{fig:Na}
\end{figure}

Figure~\ref{fig:Na} shows two of the over 700 images of the sodium
nebula recorded by IoIO and processed as described above.  Data were
recorded on over 150 nights between IoIO commissioning in March 2017
and the end of the Jovian opposition in July 2018.  \deleted{An
  animation of the IoIO sodium images, available in the on-line
  Journal, clearly show the three-dimensional structure of the
  ``banana,'' ``jet,'' and ``stream'' features discovered by
  \citet{schneider91Na} and modeled in detail by \citet{wilson02}.}
Subsequent work will address \replaced{these features}{the ``banana,''
  ``jet,'' and ``stream'' features described in the caption of Figure
  \ref{fig:Na}}.  For our current work, we concentrate on the diffuse
emission in the images, which can be studied using the surface
brightness in various apertures centered on Jupiter.  In
\S\ref{results} and Figure~\ref{fig:Na_vs_T}, we present the time
evolution of these surface brightness values to show that there was a
large modulation in the emission detected by IoIO during the 2018
Jovian opposition.  In \S\ref{discussion}, we demonstrate that this
emission was from the Jovian sodium nebula.

\section{Results}
\label{results}

\begin{figure}[h]
\centering
\plottwo{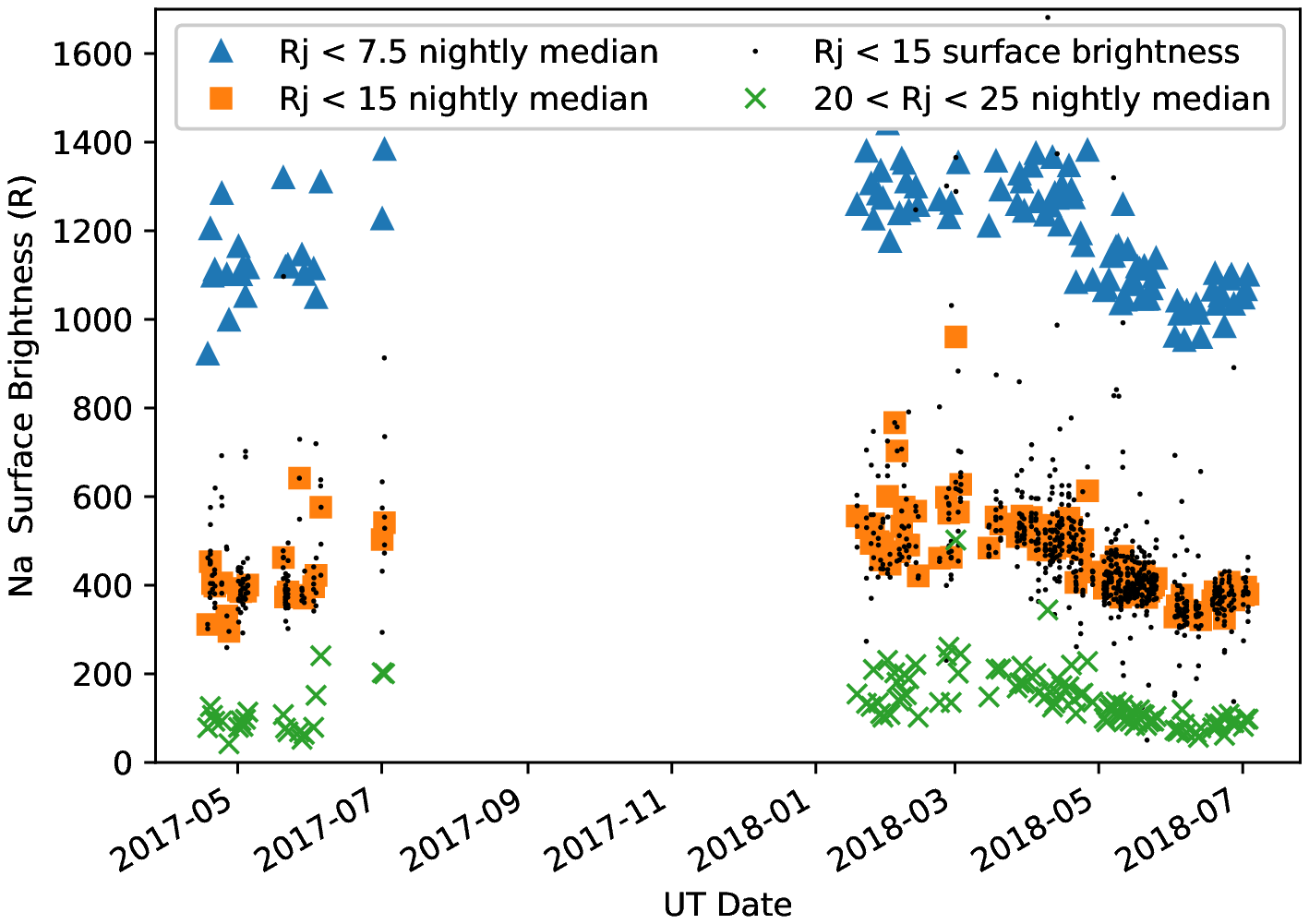}{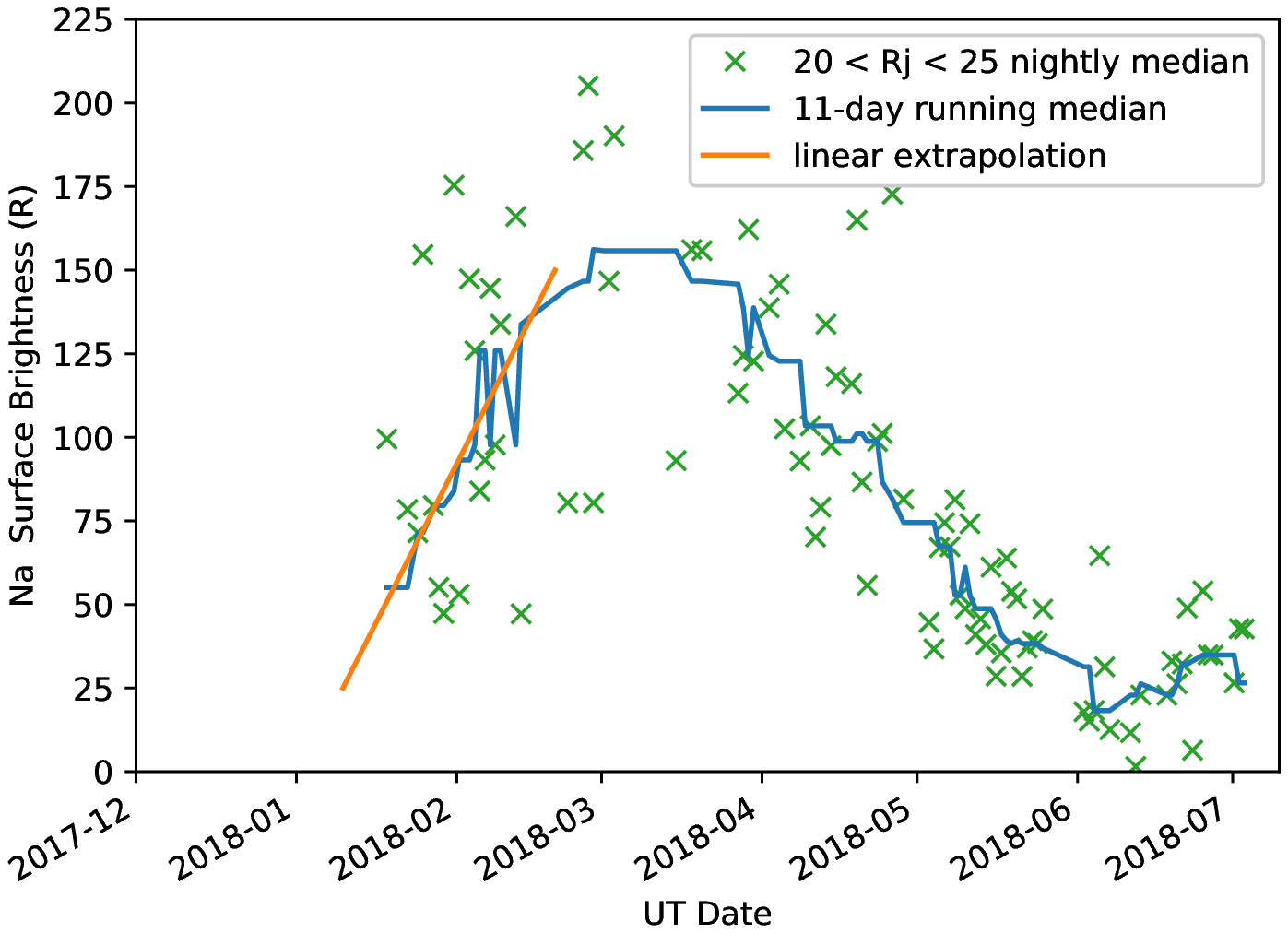}

\caption{\added{Left:} Time history of the emission measured by IoIO
  for the apertures indicated in Figure~\ref{fig:Na}.
  Section~\ref{discussion} provides the evidence that the long-term
  trend seen in these curves is long-term modulation in the Jovian
  sodium nebula surface brightness.  \added{Right: Time history of
    nightly medians from the $20\,\mathrm{R_{j}} < r <
    25\,\mathrm{R_{j}}$ aperture with estimated mesospheric sodium
    emission subtracted (\S\ref{discussion}).  An 11-day running
    median (blue histogram) is extended by a simple linear
    extrapolation (orange line) to indicate early January 2018 is the
    latest time the enhancement could have started.  The code
    \texttt{read\_ap.py} in \citet{morgenthaler19IoIO_code} reads the
    aperture sum data file in \citet{morgenthaler19_IoIO_data} to
    create both panels in the Figure.}}
\label{fig:Na_vs_T}
 \end{figure}

Figure~\ref{fig:Na_vs_T} shows that there was a significant and
long-lasting enhancement in emission detected by IoIO during the 2018
Jovian opposition.  The \added{left panel of the} Figure shows via
large colored triangles, squares and Xs, the nightly medians of the
average surface brightnesses within the three regions indicated by
colored squares in Figure~\ref{fig:Na}.  All of the surface brightness
measurements for one of the apertures are shown as small black dots.
\added{The right panel of Figure \ref{fig:Na_vs_T} shows an 11-day
  moving median (blue histogram) which smooths the effects of variable
  weather.}  \deleted{The enhancement had already begun when we
  started observation for the season, peaked in early March and
  diminished starting in late April.}  \added{As discussed in more
  detail in \S\ref{discussion},} the enhancement bears the mark of
modulation in brightness of a centrally peaked source because the more
centrally concentrated apertures have larger modulations as a function
of time.  \added{The extrapolation of the $20\,\mathrm{R_{j}} < r <
  25\,\mathrm{R_{j}}$ aperture, shown in Figure \ref{fig:Na_vs_T}
  (right, orange line), suggests that the enhancement started no later
  than early January 2018.  The scatter in the data suggests the
  enhancement could have begun as early as mid-December 2017.  The
  emission peaked in brightness in March and remained bright until
  2018 June, making this 1.8 times longer than the events observed by
  \citet{brown97} and \citet{yoneda15} and comparable in length to
  that observed by the \textsl{Galileo} Dust Detector in 2000
  \citep{krueger03}.  The intensity of the enhancement is discussed in
  more detail in \S\ref{discussion}.} \deleted{We demonstrate in
  \S\ref{discussion} that this centrally peaked source is the Jovian
  sodium nebula.}

\section{Discussion}
\label{discussion}

As discussed in \S\ref{observations}--\S\ref{reduction}, IoIO does not
see to the edge of the Jovian sodium nebula.  Furthermore, \added{to
  maximize observing time on the plasma torus, sodium} sky background
observations away from Jupiter were not systematically recorded.
Thus, we must take some care in our analyses to ensure we are
detecting modulation in the Jovian sodium nebula and not some other
source.

The first factor we consider which could possibly contribute to the
long-term modulation seen in Figure~\ref{fig:Na_vs_T} is improper
subtraction of the continuum light recorded in our on-band images.
This is particularly concerning given the systematic change in
OFFSCALE noted in \S\ref{reduction}.  We rule out this concern in
several ways.  First, the change in OFFSCALE occurred more abruptly in
April compared to the decline see in Figure~\ref{fig:Na_vs_T}.
Second, we reversed the sense of the long-term change in OFFSCALE and
re-processed images in March and June and found that the March
aperture surface brightness values were still higher than the June.
Perhaps most convincingly, we create plots like
Figure~\ref{fig:Na_vs_T} using our on-band and off-band images
separately.  These plots show more scatter than
Figure~\ref{fig:Na_vs_T}, but the on-band plot already shows the trend
seen in Figure~\ref{fig:Na_vs_T}.  The off-band plot shows no
long-term trend.  These observations confirm that our background
subtraction is reasonable and that the modulation seen in
Figure~\ref{fig:Na_vs_T} comes from line emission and not continuum
emission in the on-band filter bandpass.

Next we consider the response of IoIO to the primary source of sodium
emission other than the Jovian sodium nebula: the Earth's mesospheric
sodium layer.  This layer is formed from the ablation of meteors.  Its
thickness has seasonal dependence in the same sense as the modulation
seen in Figure~\ref{fig:Na_vs_T} \citep[e.g.,][their
  Figure~4]{dunker15}, which is why it is of concern for our analyses.
Mesospheric sodium also has a diurnal variation because it is excited
by photochemical processes local to the layer
\citep[e.g.,][]{kirchhoff79}.  In contrast to this, the Jovian sodium
nebula has negligible nightly modulation over the
$\sim3.5\times10^6$\,km region covered by the IoIO FOV.  Thus, by
considering our data on a night-by-night basis, we can probe the
response of IoIO to a uniform field-filling source without
interference from the nebula.  The black dots in
Figure~\ref{fig:Na_vs_T} \added{(left)} show the extent of the nightly
modulations for the $r < 15\,\mathrm{R_{j}}$ aperture fall within the
40\,R -- 200\,R range seen at other locations.  Larger excursions,
\added{such as the last two nights in the 2017 observing season,
  recorded as the monsoon season started,} are due to passing clouds.
Nightly variations in emission in \deleted{the} all the apertures are
highly correlated with correlation coefficients tending to one, as
expected for variation in a uniform, field filling source.  Thus, we
simultaneously confirm with the IoIO data themselves the design
criterion that IoIO's detection efficiency is flat as a function of
position (\S\ref{observations}) and that seasonal modulation of a
field-filling source would result in equal responses in all the
apertures.  We show in the next paragraph, this is not what is seen in
Figure~\ref{fig:Na_vs_T} \added{(left)}.


To demonstrate that IoIO detected modulation in the brightness of the
Jovian sodium nebula, we point out that the curve for each aperture in
Figure~\ref{fig:Na_vs_T} \added{(left)} has a unique
shape. \deleted{with} \added{Relative to their respective baselines,}
the more centrally concentrated apertures \replaced{having}{have}
larger \added{absolute} amplitudes.  This is the signature of
modulation in the brightness of a centrally peaked source.  The
baseline values for our inner ($r < 7.5\,\mathrm{R_{j}}$), middle ($r
< 15\,\mathrm{R_{j}}$), and outer ($20\,\mathrm{R_{j}} < r <
25\,\mathrm{R_{j}}$) apertures are $1030 \pm 30$\,R, $370 \pm 25$\,R,
and $80 \pm 15$\,R, respectively.  The peak amplitudes of the middle
and inner apertures are factors of $\sim$1.5 and $\sim$ 2.2 higher
than the outer aperture, respectively.  After removal of their
respective baselines and scaling, the curves from the three apertures
are in good agreement.  Were we seeing seasonal modulation in the
telluric sodium layer, the amplitudes would all have the same values
in the same way \deleted{they do to nightly modulation in the telluric
  sodium.}\added{that the nightly modulations do.}  This is our most
convincing evidence that we are detecting modulation in the Jovian
sodium nebula.

Although we have ruled out mesospheric emission as the cause of the
long-term modulation seen in Figure~\ref{fig:Na_vs_T} \added{(left)},
we cannot rule out its contribution as a relatively stable background.
In fact, we expect it.  As discussed above, we did not record
systematic sky background measurements, so this is not something that
we can estimate independently.  Instead, we compare the measured
surface brightness in our outer aperture during the nebula's quiescent
state to the surface brightness of inner aperture used by
\citet{yoneda09} during similarly quiet conditions.  Both these
apertures correspond to $r \sim 25\,\mathrm{R_{j}}$.  As quoted above,
the baseline in our outer aperture is $80 \pm 15$\,R.  The
mesosphere-subtracted value quoted by \citet{yoneda09} is 20\,R --
30\,R.  This suggests that 50\,R -- 60\,R of our emission is
mesospheric, which is comparable to baseline values measured by this
team at other locations.  \deleted{Subtracting that from the $225 \pm
  15$\,R peak observed in our $r \sim 25\,\mathrm{R_{j}}$ aperture in
  late February and early March suggests that the amplitude of the
  modulation observed in the Jovian sodium nebula is 200$\pm$20\,R or
  a factor of more than two larger than the 70 -- 80\,R peak in the
  event measured by \citet{yoneda09}.}  \added{Subtracting this from
  our $r \sim 25\,\mathrm{R_{j}}$ aperture results in
  Figure~\ref{fig:Na_vs_T} (right), which shows that the peak
  amplitude of the modulation observed in the Jovian sodium nebula at
  $r \sim 25\,\mathrm{R_{j}}$ is 155$\pm$25\,R or a factor of $\sim$2
  larger than the 70 -- 80\,R peak in the event measured by
  \citet{yoneda09}.  The peak amplitude in the 2015 January to 2015
  April enhancement reported by \citet{yoneda15} was a factor of
  $\sim$1.5 larger than the \citet{yoneda09} enhancement.  Thus, the
  2018 enhancement detected by IoIO was a factor of $\sim$1.3 brighter
  than the 2015 January to 2015 April enhancement reported by
  \citet{yoneda15}.}

\section{Conclusion}
\label{conclusion}

\deleted{Following the discussion in \S\ref{intro}, we associate the large
modulation in the surface brightness of the Jovian sodium nebula
presented in \S\ref{results} and Figure~\ref{fig:Na_vs_T} with a
volcanic event on Io.  A rough extrapolation of the data suggests that
the event occurred some time between mid December 2017 and early
January 2018.  The calculations detailed in \S\ref{discussion}
suggests the event was more than twice as bright as that reported by
\citet{yoneda09}.  The nebula remained bright until 2018 June, making
this the longest event observed by this technique and comparable in
length to that observed by the \textsl{Galileo} Dust Detector in 2000
\citep{krueger03}.}

\added{We have detected a large and long-lasting enhancement in the
  Jovian sodium nebula.  The extrapolation of the data shown in Figure
  \ref{fig:Na_vs_T} (right) suggests that the enhancement started no
  later than early January 2018.  The scatter in the data suggests the
  enhancement could have begun as early as mid-December 2017.  The
  calculations detailed in \S\ref{discussion} suggests the event was
  $\sim$30\% brighter than the primary enhancement reported by
  \citet{yoneda15}.  The nebula remained bright until 2018 June,
  making this 1.8 times longer than the events observed by
  \citet{brown97} and \citet{yoneda15} and comparable in length to
  that observed by the \textsl{Galileo} Dust Detector in 2000
  \citep{krueger03}.}

\deleted{Infrared observations available to this team, recorded at
  NASA's IRTF between 2017-12-16 and 2018-01-15, do not detect any
  unusual activity during this time period, with the caveat that
  longitude ranges $120^{\circ}\sim210^{\circ}$ were not observed.  No
  other report of notable volcanic events before May 2018 have been
  made by infrared astronomers, despite the fact that a substantial
  number of observations of were conducted in support of NASA's
  \textsl{Juno} mission.  This is suggestive, but not conclusive
  evidence that the type of event observed by IoIO may be difficult to
  detect with current ground-based infrared telescopic observation
  techniques.}

\added{Infrared observations recorded at NASA's IRTF by our team and
  at the W.\ M.\ Keck Observatory by K.\ de Kleer \& I.\ de Pater as a
  continuation of the monitoring program discussed in
  \citet{deKleer16} achieved full longitudinal coverage of Io, but
  detected no IR-bright eruptions that lasted for over a month in the
  December 2017 to January 2018 timeframe.  A new eruption near
  Susanoo/Mulungu paterae (20\arcdeg N 218\arcdeg W) was observed to
  begin sometime between January 2 and 12 (personal communication, de
  Kleer \& de Pater, Dec.\ 2018).  This event was
  20\,GW\,$\mu$m$^{-1}$sr$^{-1}$ at brightest detected Lp (3.78
  $\mu$m) and would therefore be classified as a faint eruption in the
  taxonomic scheme of \citet{deKleer16}.  The proximity in time
  between this eruption and the onset of the sodium nebula enhancement
  is suggestive but not conclusive evidence of a relationship.  A more
  convincing case was made by \citet{deKleer16} that the two sodium
  nebula enhancements seen by \citet{yoneda15} were associated with
  two eruptions in the ``mini-outburst'' class at Kurdalagon Patera,
  as both eruptions were contemporaneous with the onset of the nebula
  enhancements.  However, as noted by \citet*{deKleer16AGU}, during
  the three-year study of \citet{deKleer16}, not all of the detected
  sodium nebula enhancements had identifiable IR counterparts.  Along
  the same lines, the two Kurdalagon Patera outbursts were a factor of
  $\sim$3 brighter than the eruption near Susanoo/Mulungu paterae, yet
  the two enhancements found by \citet{yoneda15} were not of
  equivalent size nor were they larger than the enhancement reported
  here, as one would expect if IR brightness was correlated with the
  amount of gas released.  The picture that emerges is that IR
  activity on Io is simply not predictive of sodium nebula
  enhancement.  This observation is strengthened by the fact that all
  of the associations between IR eruptions and sodium nebula
  enhancements have been made \textit{a posteriori}.  This highlights
  the importance of synoptic observations of the Jovian sodium nebula
  for monitoring the supply of material to Jupiter's magnetosphere --
  material that drives a host of magnetospheric phenomena.}

With our detection of such a long-lasting event during the first half
of the 2018 Jovian opposition, other observations may be placed in
context.  This is particularly important for observations conducted by
\textsl{Juno}, large-aperture observatories, and \textsl{HST}, which
themselves do not have synoptic coverage comparable to IoIO and were
therefore not able to independently detect this event.  For instance,
our team regularly conducts observations of the IPT with the ARC
3.5\,m telescope at Apache Point Observatory \citep{schmidt18}.  In
May 2018, these were seen to be the brightest \added{yet} recorded by
this facility.  Preliminary reduction of our IoIO [\ion{S}{2}] images
also shows evidence that the overall brightness of the IPT follows a
similar envelope to that observed by \citet{brown97} during the
volcanic event they saw (\S\ref{intro}).  EUV observations of the IPT
by \textsl{Hisaki} over this time period should be brighter than
normal and show chemical and periodicity changes similar to those seen
by \citet{steffl08} and \citet{kimura18} after volcanic events.  We
predict the neutral oxygen cloud around Jupiter, detectable with the
\textsl{Hisaki} satellite \citep{koga18IC, koga18JGR}, will show
higher values than found previously.  Finally, higher than average
auroral activity on Jupiter should be detected by \textit{in situ}
measurements from the \textsl{JUNO/JADE} instrument; in the UV by
\textsl{Juno/UVS}, \textsl{Hisaki}, and \textsl{HST}; in the infrared
by \textsl{Juno/JIRAM} and ground-based infrared telescopes; and in
the radio by \textsl{Juno/WAVES} and ground-based radio telescopes
such as the Nan\c{c}ay Decametric Array \citep[e.g.,][]{mccomas17,
  gladstone17, kimura15, kita16, kurth17, radioti13, marques17}.

\added{Although we cannot provide independent measurement of the
  geological and atmospheric processes responsible for the production
  and release of gas detected from Io, we can use the shape of the
  11-day running median in Figure \ref{fig:Na_vs_T} (right) to provide
  an estimate of the time evolution of the gas release, which, upon
  further study, may provide clues to its origin.  Electron impact
  ionization is the primary loss mechanism of sodium within the IoIO
  FOV and is of order 10 -- 20 days \citep[e.g.,][their Figure
    3]{wilson02}, which is short compared to the $\sim$180 day
  enhancement in the nebula.  Thus, the shape of the 11-day running
  median is primarily determined by the physical processes responsible
  for gas release from Io.}

\acknowledgments

\deleted{The Python code used to control the telescope and reduce the
  data is available at \url{http://github.com/jpmorgen/IoIO}.  Code
  used for debugging this suite, created by Daniel R.\ Morgenthaler,
  is found at \url{http://github.com/jpmorgen/Daniel}.  Long-term, an
  on-going archive of all raw and reduced images collected by IoIO
  will be kept at NASA's Planetary Data System (PDS).}

We thank Scott Tucker of Starizona for his excellent mechanical design
and construction of the IoIO coronagraph and Vishnu Ready for helping
to make that contact.  We also acknowledge Dean Salman, manager of the
San Pedro Valley Observatory, hosting site of IoIO, whose expertise at
small-aperture astronomy was a major contributing factor to our
ability to record scientifically useful images two weeks after the
receipt of the last major and longest lead time parts (the filters).
This work is supported by NSF grant AST 1616928 to the Planetary
Science Institute.

\software{AstroPy \citep{astropy_collaboration13},
  Astroquery \citep{astroquery13},
  ccdproc \citep{ccdproc},
  \added{Python aliases and shortcuts \citep{morgenthaler_d19}},
  DataThief~III \citep{tummers06},
  IDL \citep{IDL},
  \added{IoIO control, reduction, and analysis software
  \citep{morgenthaler19IoIO_code}},
  matplotlib \citep{matplotlib},
  NumPy \citep{oliphant06},
  scikit-image \citep{scikit-image}
}





\bibliography{cv-jour, io,comets,cross_sections,Fabry_Perot,dxs,xqc,thesis,SHS,UV_Back,sun,asteroids,software,atmosphere}

\begin{thebibliography}{}
\expandafter\ifx\csname natexlab\endcsname\relax\def\natexlab#1{#1}\fi
\providecommand{\url}[1]{\href{#1}{#1}}
\providecommand{\dodoi}[1]{doi:~\href{http://doi.org/#1}{\nolinkurl{#1}}}
\providecommand{\doeprint}[1]{\href{http://ascl.net/#1}{\nolinkurl{http://ascl.net/#1}}}
\providecommand{\doarXiv}[1]{\href{https://arxiv.org/abs/#1}{\nolinkurl{https://arxiv.org/abs/#1}}}

\bibitem[{{Astropy Collaboration} {et~al.}(2013){Astropy Collaboration},
  {Robitaille}, {Tollerud}, {Greenfield}, {Droettboom}, {Bray}, {Aldcroft},
  {Davis}, {Ginsburg}, {Price-Whelan}, {Kerzendorf}, {Conley}, {Crighton},
  {Barbary}, {Muna}, {Ferguson}, {Grollier}, {Parikh}, {Nair}, {Unther},
  {Deil}, {Woillez}, {Conseil}, {Kramer}, {Turner}, {Singer}, {Fox}, {Weaver},
  {Zabalza}, {Edwards}, {Azalee Bostroem}, {Burke}, {Casey}, {Crawford},
  {Dencheva}, {Ely}, {Jenness}, {Labrie}, {Lim}, {Pierfederici}, {Pontzen},
  {Ptak}, {Refsdal}, {Servillat}, \& {Streicher}}]{astropy_collaboration13}
{Astropy Collaboration}, {Robitaille}, T.~P., {Tollerud}, E.~J., {et~al.} 2013,
  Astron. Astrophys., 558, A33, \dodoi{10.1051/0004-6361/201322068}

\bibitem[{Brown \& Bouchez(1997)}]{brown97}
Brown, M.~E., \& Bouchez, A.~H. 1997, Sci, 278, 268,
  \dodoi{10.1126/science.278.5336.268}

\bibitem[{Brown \& Chaffee(1974)}]{brown74}
Brown, R.~A., \& Chaffee, Jr., F.~H. 1974, Astrophys. J., Lett., 187, L125,
  \dodoi{10.1086/181413}

\bibitem[{Craig {et~al.}(2015)Craig, Crawford, Deil, Gomez, G{\"u}nther, Heidt,
  Horton, Karr, Nelson, Ninan, Pattnaik, Rol, Schoenell, Seifert, Singh,
  Sipocz, Stotts, Streicher, Tollerud, Walker, \& {ccdproc
  contributors}}]{ccdproc}
Craig, M.~W., Crawford, S.~M., Deil, C., {et~al.} 2015, {ccdproc: CCD data
  reduction software}, Astrophysics Source Code Library.
\newblock \doeprint{1510.007}

\bibitem[{{de Kleer} \& {de Pater}(2016)}]{deKleer16}
{de Kleer}, K., \& {de Pater}, I. 2016, Icarus, 280, 378,
  \dodoi{10.1016/j.icarus.2016.06.019}

\bibitem[{{de Kleer} {et~al.}(2016){de Kleer}, {de Pater}, \&
  Yoneda}]{deKleer16AGU}
{de Kleer}, K., {de Pater}, I., \& Yoneda, M. 2016, AGU Fall Meeting Abstracts,
  P21E

\bibitem[{{de Pater} {et~al.}(2017){de Pater}, {de Kleer}, Davies, \&
  {\'A}d{\'a}mkovics}]{depater17}
{de Pater}, I., {de Kleer}, K., Davies, A.~G., \& {\'A}d{\'a}mkovics, M. 2017,
  Icarus, 297, 265, \dodoi{10.1016/j.icarus.2017.03.016}

\bibitem[{Dunker {et~al.}(2015)Dunker, Hoppe, Feng, Plane, \& Marsh}]{dunker15}
Dunker, T., Hoppe, U.-P., Feng, W., Plane, J.~M.~C., \& Marsh, D.~R. 2015,
  Journal of Atmospheric and Solar-Terrestrial Physics, 127, 111,
  \dodoi{10.1016/j.jastp.2015.01.003}

\bibitem[{Ginsburg {et~al.}(2013)Ginsburg, Robitaille, Parikh, Deil, Mirocha,
  Woillez, Svoboda, Willett, T~Allen, Grollier, \& et~al.}]{astroquery13}
Ginsburg, A., Robitaille, T., Parikh, M., {et~al.} 2013, Astroquery v0.1,
  figshare, \dodoi{10.6084/m9.figshare.805208.v2}.
\newblock \url{https://figshare.com/articles/Astroquery_v0_1/805208/2}

\bibitem[{Gladstone {et~al.}(2017)Gladstone, Persyn, Eterno, Walther, Slater,
  Davis, Versteeg, Persson, Young, Dirks, Sawka, Tumlinson, Sykes, Beshears,
  Rhoad, Cravens, Winters, Klar, Lockhart, Piepgrass, Greathouse, Trantham,
  Wilcox, Jackson, Siegmund, Vallerga, Raffanti, Martin, G{\'e}rard, Grodent,
  Bonfond, Marquet, \& Denis}]{gladstone17}
Gladstone, G.~R., Persyn, S.~C., Eterno, J.~S., {et~al.} 2017, Space Science
  Reviews, 213, 447, \dodoi{10.1007/s11214-014-0040-z}

\bibitem[{{Harris Geospatial Solutions}(2018)}]{IDL}
{Harris Geospatial Solutions}. 2018, {The Interactive Data Language}.
\newblock \url{https://www.harrisgeospatial.com}

\bibitem[{Hunter(2007)}]{matplotlib}
Hunter, J.~D. 2007, Computing In Science \& Engineering, 9, 90,
  \dodoi{10.1109/MCSE.2007.55}

\bibitem[{Jessup \& Spencer(2012)}]{jessup12}
Jessup, K.~L., \& Spencer, J.~R. 2012, Icarus, 218, 378,
  \dodoi{10.1016/j.icarus.2011.11.013}

\bibitem[{Johnson {et~al.}(1995)Johnson, Matson, Blaney, Veeder, \&
  Davies}]{johnson95}
Johnson, T.~V., Matson, D.~L., Blaney, D.~L., Veeder, G.~J., \& Davies, A.
  1995, Geophys. Res. Lett., 22, 3293, \dodoi{10.1029/95GL03084}

\bibitem[{Karkoschka(1998)}]{karkoschka98}
Karkoschka, E. 1998, Icarus, 133, 134, \dodoi{10.1006/icar.1998.5913}

\bibitem[{Kimura {et~al.}(2015)Kimura, Badman, Tao, Yoshioka, Murakami,
  Yamazaki, Tsuchiya, Bonfond, Steffl, Masters, Kasahara, Hasegawa, Yoshikawa,
  Fujimoto, \& Clarke}]{kimura15}
Kimura, T., Badman, S.~V., Tao, C., {et~al.} 2015, Geophys. Res. Lett., 42,
  1662, \dodoi{10.1002/2015GL063272}

\bibitem[{Kimura {et~al.}(2018)Kimura, Hiraki, Tao, Tsuchiya, Delamere,
  Yoshioka, Murakami, Yamazaki, Kita, Badman, Fukazawa, Yoshikawa, \&
  Fujimoto}]{kimura18}
Kimura, T., Hiraki, Y., Tao, C., {et~al.} 2018, J. Geophys. Res., 123, 1885,
  \dodoi{10.1002/2017JA025029}

\bibitem[{Kirchhoff {et~al.}(1979)Kirchhoff, Clemesha, \&
  Simonich}]{kirchhoff79}
Kirchhoff, V.~W.~J.~H., Clemesha, B.~R., \& Simonich, D.~M. 1979, J. Geophys.
  Res., 84, 1323, \dodoi{10.1029/JA084iA04p01323}

\bibitem[{Kita {et~al.}(2016)Kita, Kimura, Tao, Tsuchiya, Misawa, Sakanoi,
  Kasaba, Murakami, Yoshioka, Yamazaki, Yoshikawa, \& Fujimoto}]{kita16}
Kita, H., Kimura, T., Tao, C., {et~al.} 2016, Geophys. Res. Lett., 43, 6790,
  \dodoi{10.1002/2016GL069481}

\bibitem[{Kliore {et~al.}(1975)Kliore, Fjeldbo, Seidel, Sweetnam, Sesplaukis,
  Woiceshyn, \& Rasool}]{kliore75}
Kliore, A.~J., Fjeldbo, G., Seidel, B.~L., {et~al.} 1975, Icarus, 24, 407,
  \dodoi{10.1016/0019-1035(75)90057-3}

\bibitem[{Koga {et~al.}(2018{\natexlab{a}})Koga, Tsuchiya, Kagitani, Sakanoi,
  Yoneda, Yoshioka, Kimura, Murakami, Yamazaki, Yoshikawa, \& Smith}]{koga18IC}
Koga, R., Tsuchiya, F., Kagitani, M., {et~al.} 2018{\natexlab{a}}, Icarus, 299,
  300, \dodoi{10.1016/j.icarus.2017.07.024}

\bibitem[{Koga {et~al.}(2018{\natexlab{b}})Koga, Tsuchiya, Kagitani, Sakanoi,
  Yoneda, Yoshioka, Yoshikawa, Kimura, Murakami, Yamazaki, Smith, \&
  Bagenal}]{koga18JGR}
---. 2018{\natexlab{b}}, J. Geophys. Res., 123, 3764,
  \dodoi{10.1029/2018JA025328}

\bibitem[{Kr{\"u}ger {et~al.}(2003)Kr{\"u}ger, Geissler, Hor{\'a}nyi, Graps,
  Kempf, Srama, Moragas-Klostermeyer, Moissl, Johnson, \& Gr{\"u}n}]{krueger03}
Kr{\"u}ger, H., Geissler, P., Hor{\'a}nyi, M., {et~al.} 2003, Geophys. Res.
  Lett., 30, 210000, \dodoi{10.1029/2003GL017827}

\bibitem[{Kupo {et~al.}(1976)Kupo, Mekler, \& Eviatar}]{kupo76}
Kupo, I., Mekler, Y., \& Eviatar, A. 1976, Astrophys. J., Lett., 205, L51,
  \dodoi{10.1086/182088}

\bibitem[{Kurth {et~al.}(2017)Kurth, Hospodarsky, Kirchner, Mokrzycki,
  Averkamp, Robison, Piker, Sampl, \& Zarka}]{kurth17}
Kurth, W.~S., Hospodarsky, G.~B., Kirchner, D.~L., {et~al.} 2017, Space Science
  Reviews, 213, 347, \dodoi{10.1007/s11214-017-0396-y}

\bibitem[{Kurucz(2005)}]{kurucz05ATLAS}
Kurucz, R.~L. 2005, Memorie della Societa Astronomica Italiana Supplement, 8,
  14

\bibitem[{Marques {et~al.}(2017)Marques, Zarka, Echer, Ryabov, Alves, Denis, \&
  Coffre}]{marques17}
Marques, M.~S., Zarka, P., Echer, E., {et~al.} 2017, Astron. Astrophys., 604,
  A17, \dodoi{10.1051/0004-6361/201630025}

\bibitem[{McComas {et~al.}(2017)McComas, Alexander, Allegrini, Bagenal, Beebe,
  Clark, Crary, Desai, De Los~Santos, Demkee, Dickinson, Everett, Finley,
  Gribanova, Hill, Johnson, Kofoed, Loeffler, Louarn, Maple, Mills, Pollock,
  Reno, Rodriguez, Rouzaud, Santos-Costa, Valek, Weidner, Wilson, Wilson, \&
  White}]{mccomas17}
McComas, D.~J., Alexander, N., Allegrini, F., {et~al.} 2017, Space Science
  Reviews, 213, 547, \dodoi{10.1007/s11214-013-9990-9}

\bibitem[{McGrath {et~al.}(2004)McGrath, Lellouch, Strobel, Feldman, \&
  Johnson}]{mcgrath04}
McGrath, M.~A., Lellouch, E., Strobel, D.~F., Feldman, P.~D., \& Johnson, R.~E.
  2004, {Satellite atmospheres}, ed. F.~Bagenal, T.~E. Dowling, \& W.~B.
  {McKinnon} (Cambridge: Cambridge University Press), 457--483

\bibitem[{Mendillo {et~al.}(1990)Mendillo, Baumgardner, Flynn, \&
  Hughes}]{mendillo90}
Mendillo, M., Baumgardner, J., Flynn, B., \& Hughes, W.~J. 1990, Nature, 348,
  312, \dodoi{10.1038/348312a0}

\bibitem[{Mendillo {et~al.}(2004)Mendillo, Wilson, Spencer, \&
  Stansberry}]{mendillo04}
Mendillo, M., Wilson, J., Spencer, J., \& Stansberry, J. 2004, Icarus, 170,
  430, \dodoi{10.1016/j.icarus.2004.03.009}

\bibitem[{Morabito {et~al.}(1979)Morabito, Synnott, Kupferman, \&
  Collins}]{morabito79}
Morabito, L.~A., Synnott, S.~P., Kupferman, P.~N., \& Collins, S.~A. 1979, Sci,
  204, 972, \dodoi{10.1126/science.204.4396.972}

\bibitem[{Morgenthaler \& Morgenthaler(2019)}]{morgenthaler_d19}
Morgenthaler, D.~R., \& Morgenthaler, J.~P. 2019, Python aliases and shortcuts,
  v1.0.1,  Zenodo, \dodoi{10.5281/zenodo.2535735}.
\newblock \url{https://doi.org/10.5281/zenodo.2535735}

\bibitem[{Morgenthaler(2019)}]{morgenthaler19IoIO_code}
Morgenthaler, J.~P. 2019, {Io Input/Output facility (IoIO) control, reduction,
  and analysis software}, v1.0.1,  Zenodo, \dodoi{10.5281/zenodo.2535838}.
\newblock \url{https://doi.org/10.5281/zenodo.2535838}

\bibitem[{Morgenthaler {et~al.}(2019)Morgenthaler, Rathbun, Schmidt,
  Baumgardner, \& Schneider}]{morgenthaler19_IoIO_data}
Morgenthaler, J.~P., Rathbun, J.~A., Schmidt, C.~A., Baumgardner, J., \&
  Schneider, N.~M. 2019, Io Input/Output (IoIO) aperture sum data file, v1.0,
  Zenodo, \dodoi{10.5281/zenodo.2535920}.
\newblock \url{https://doi.org/10.5281/zenodo.2535920}

\bibitem[{Nozawa {et~al.}(2004)Nozawa, Misawa, Takahashi, Morioka, Okano, \&
  Sood}]{nozawa04_no_SII}
Nozawa, H., Misawa, H., Takahashi, S., {et~al.} 2004, J. Geophys. Res., 109,
  7209, \dodoi{10.1029/2003JA010241}

\bibitem[{Oliphant(2006)}]{oliphant06}
Oliphant, T.~E. 2006, {Guide to NumPy} (Provo: Trelgol Publishing)

\bibitem[{Radioti {et~al.}(2013)Radioti, Lystrup, Bonfond, Grodent, \&
  G{\'e}rard}]{radioti13}
Radioti, A., Lystrup, M., Bonfond, B., Grodent, D., \& G{\'e}rard, J.-C. 2013,
  J. Geophys. Res., 118, 2286, \dodoi{10.1002/jgra.50245}

\bibitem[{Rathbun \& Spencer(2006)}]{rathbun06}
Rathbun, J.~A., \& Spencer, J.~R. 2006, Geophys. Res. Lett., 33, L17201,
  \dodoi{10.1029/2006GL026844}

\bibitem[{Schmidt {et~al.}(2018)Schmidt, Schneider, Leblanc, Gray,
  Morgenthaler, Turner, \& Grava}]{schmidt18}
Schmidt, C., Schneider, N., Leblanc, F., {et~al.} 2018, J. Geophys. Res., 123,
  5610, \dodoi{10.1029/2018JA025296}

\bibitem[{Schneider \& Bagenal(2007)}]{schneider07}
Schneider, N.~M., \& Bagenal, F. 2007, {Io's neutral clouds, plasma torus, and
  magnetospheric interaction} (Berlin, Heidelberg: Springer Praxis Books /
  Geophysical Sciences), 265--286

\bibitem[{Schneider {et~al.}(1991)Schneider, Wilson, Trauger, Brown, Evans, \&
  Shemansky}]{schneider91Na}
Schneider, N.~M., Wilson, J.~K., Trauger, J.~T., {et~al.} 1991, Sci, 253, 1394,
  \dodoi{10.1126/science.253.5026.1394}

\bibitem[{Sinton(1980)}]{sinton80b}
Sinton, W.~M. 1980, Icarus, 43, 56, \dodoi{10.1016/0019-1035(80)90087-1}

\bibitem[{Steffl {et~al.}(2006)Steffl, Delamere, \& Bagenal}]{steffl06}
Steffl, A.~J., Delamere, P.~A., \& Bagenal, F. 2006, Icarus, 180, 124,
  \dodoi{10.1016/j.icarus.2005.07.013}

\bibitem[{Steffl {et~al.}(2008)Steffl, Delamere, \& Bagenal}]{steffl08}
---. 2008, Icarus, 194, 153, \dodoi{10.1016/j.icarus.2007.09.019}

\bibitem[{Tummers(2006)}]{tummers06}
Tummers, B. 2006, {DataThief III}.
\newblock \url{https://datathief.org/}

\bibitem[{{van der Walt} {et~al.}(2014){van der Walt}, Sch{\"o}nberger,
  Nunez-Iglesias, Boulogne, Warner, Yager, Gouillart, \& Yu}]{scikit-image}
{van der Walt}, S., Sch{\"o}nberger, J.~L., Nunez-Iglesias, J., {et~al.} 2014,
  PeerJ, 2, e453, \dodoi{10.7717/peerj.453}

\bibitem[{Wilson {et~al.}(2002)Wilson, Mendillo, Baumgardner, Schneider,
  Trauger, \& Flynn}]{wilson02}
Wilson, J.~K., Mendillo, M., Baumgardner, J., {et~al.} 2002, Icarus, 157, 476,
  \dodoi{10.1006/icar.2002.6821}

\bibitem[{Witteborn {et~al.}(1979)Witteborn, Bregman, \& Pollack}]{witteborn79}
Witteborn, F.~E., Bregman, J.~D., \& Pollack, J.~B. 1979, Sci, 203, 643,
  \dodoi{10.1126/science.203.4381.643}

\bibitem[{Woodman {et~al.}(1979)Woodman, Cochran, \& Slavsky}]{woodman79}
Woodman, J.~H., Cochran, W.~D., \& Slavsky, D.~B. 1979, Icarus, 37, 73,
  \dodoi{10.1016/0019-1035(79)90116-7}

\bibitem[{Yoneda {et~al.}(2009)Yoneda, Kagitani, \& Okano}]{yoneda09}
Yoneda, M., Kagitani, M., \& Okano, S. 2009, Icarus, 204, 589,
  \dodoi{10.1016/j.icarus.2009.07.023}

\bibitem[{Yoneda {et~al.}(2015)Yoneda, Kagitani, Tsuchiya, Sakanoi, \&
  Okano}]{yoneda15}
Yoneda, M., Kagitani, M., Tsuchiya, F., Sakanoi, T., \& Okano, S. 2015, Icarus,
  261, 31, \dodoi{10.1016/j.icarus.2015.07.037}

\bibitem[{Yoneda {et~al.}(2010)Yoneda, Nozawa, Misawa, Kagitani, \&
  Okano}]{yoneda10}
Yoneda, M., Nozawa, H., Misawa, H., Kagitani, M., \& Okano, S. 2010, Geophys.
  Res. Lett., 37, 11202, \dodoi{10.1029/2010GL043656}

\bibitem[{Yoneda {et~al.}(2014)Yoneda, Miyata, Tsang, Sako, Kamizuka, Nakamura,
  Asano, Uchiyama, Okada, Hayashi, Yoshii, Kagitani, Sakanoi, Kasaba, \&
  Okano}]{yoneda14}
Yoneda, M., Miyata, T., Tsang, C.~C.~C., {et~al.} 2014, Icarus, 236, 153,
  \dodoi{10.1016/j.icarus.2014.01.019}

\end{thebibliography}


\listofchanges

\end{document}